\documentstyle[11pt,epsfig]{article}
\begin{document}

\noindent{\Large{\bf Ray dynamics in ocean acoustics}}\\

\noindent{\Large{Michael G. Brown}}\\
{\footnotesize{\em Rosenstiel School of Marine and Atmospheric Science,
University of Miami, Miami, Florida 33149}}\\

\noindent{\Large{John A. Colosi}}\\
{\footnotesize{\em Woods Hole Oceanographic Institution, Woods Hole,
Massachusetts 02543}}\\

\noindent{\Large{Steven Tomsovic}}\\
{\footnotesize{\em Department of Physics, Washington State University, Pullman,
Washington 99164}}\\

\noindent{\Large{Anatoly L. Virovlyansky}}\\
{\footnotesize{\em Institute of Applied Physics, Russian Academy of Science,
6003600 Nizhny Novgorod, Russia}}\\ 

\noindent{\Large{Michael A. Wolfson}}\\
{\footnotesize{\em Applied Physics Laboratory, University of Washington,
Seattle, Washington 91805}}\\

\noindent{\Large{George M. Zaslavsky}}\\
{\footnotesize{\em Courant Institute of Mathematical Sciences, New York
University, New York, New York 10012}}\\

\noindent
Recent results relating to ray dynamics in ocean acoustics are reviewed.
Attention is focussed on long-range propagation in deep ocean
environments. For this class of problems, the ray equations may be
simplified by making use of a one-way formulation in which the range
variable appears as the independent (time-like) variable.  Topics
discussed include integrable and nonintegrable ray systems, action-angle
variables, nonlinear resonances and the KAM theorem, ray chaos, Lyapunov
exponents, predictability, nondegeneracy violation, ray intensity
statistics, semiclassical breakdown, wave chaos, and the connection
between ray chaos and mode coupling.  The Hamiltonian structure of the
ray equations plays an important role in all of these topics.  

\section*{\large INTRODUCTION}

The chaotic dynamics of ray trajectories in ocean acoustics have been
explored in a number of recent publications (\cite{AZ85}-\cite{WTo}).  The
purpose of the present paper is to provide a review of results relating to
this topic.  Our exposition is brief but is intended to be self-contained.
We introduce a sequence of ray-based simplifications to the mathematical
description of underwater sound propagation in order to get a more
complete and clear understanding of the underlying propagation physics,
especially in range-dependent environments.  We consider these
simplifications as the starting point of developing a quantitative
theory.  It is our opinion that even full wave simulations cannot be used
effectively without some understanding of the material described in this
paper.

To make our discussion more concrete, we focus our attention on long-range
propagation in deep ocean conditions.  Theoretical results are emphasized,
but with an eye toward analyzing measurements.  For this reason
considerable attention is paid to results that can be applied in the presence
of complicated (nonperiodic) range-dependent ocean structure.  In a separate
paper many of the results presented and discussed here will be applied to the
analysis of a particular data set. 

In the next section we review important preliminary material.  First, we
introduce ray-based solutions to the Helmholtz equation.  We then
introduce the one-way form of the ray equations and the standard
parabolic approximation.  Finally, the motion of rays in a
range-independent environment is discussed.  For this class of problems
the ray equations are integrable and the ray trajectories are most
naturally described using action-angle variables.  In anticipation of the
material that follows, the action-angle formalism is introduced. 

Section II focuses on the behavior of rays in range-dependent environments,
i.e., on nonintegrable ray systems.  The action-angle formalism is used here
to introduce nonlinear resonances and the KAM theorem.  The notion of ray
chaos is discussed, as are Lyapunov exponents.  Our discussion of the (well
known) limitations on the predictability of isolated chaotic trajectories is
complemented by a discussion of the (generally unappreciated) stability of
families of chaotic trajectories.  Also in this section we discuss the
connection between an important condition, known as nondegeneracy, and ray
stability.

In section III, we discuss ray intensity statistics and related topics,
including the distribution of finite range estimates of Lyapunov
exponents.  The results presented here were first described in the analysis
of an idealized underwater acoustic problem, but have since been encountered
in other applications.  Problems associated with the important task of
connecting ray intensity statistics to finite frequency wavefield
intensity statistics are discussed.

In section IV, we provide a more general, but brief, discussion of `wave
chaos' -- the study of wave systems that, in the ray limit, exhibit
chaotic motion. This topic falls slightly outside the bounds of providing
a review of ray dynamics, but is too important to omit.  Strong
results relating to this topic are dificult to obtain.  Much of our
discussion focusses on the question of whether semiclassical (ray-based)
wavefield representations break down at the so-called Ehrenfest range,
which scales as $\ln(\bar{f})$ where
$\bar{f}$ is the appropriately nondimensionalized wave frequency.

In section V, we describe the connection between ray chaos and mode
coupling. This work builds on well-known results on ray-mode duality.
The results described here provide a promising means of attacking the
wave chaos problem inasmuch as finite frequency effects are built into
the modal description of the wavefield.

In the final section, we briefly discuss two issues.  First, we discuss
the principal shortcoming of our current knowledge -- our relatively poor
understanding of the wave chaos problem.  Second, we discuss the manner in
which ideas relating to deterministic chaos complement and/or conflict with
more traditional ideas relating to the study of wave propagation in random
media. 

\section*{\large I. PRELIMINARY RESULTS}

This section provides background material that is necessary to understand the
material that is presented in the sections that follow.  Starting with the
Helmholtz equation we introduce the ray equations and their one-way form, the
standard parabolic approximation, and the action-angle description of ray
motion in range-independent environments.

\section*{\large A. Waves and rays}

Fixed-frequency (cw) acoustic wavefields satisfy the Helmholtz equation,
\begin{equation}
\label{Helm}
\nabla^2 u + \sigma^2 c^{-2}({\bf r}) u = 0,
\end{equation}
where $\sigma = 2 \pi f$ is the angular frequency of the wavefield and
$c(\bf r)$ is the sound speed.  In anticipation of our later focus on
describing sound fields generated by a point source in environments in which
focusing in the azimuthal direction is assumed to be negligible, we shall
assume propagation in two space dimensions with ${\bf r} = (z,r)$ where $z$
is depth and $r$ is range.  The so-called short wave approximation can be
used when
\begin{equation}
\sigma \gg \nabla c,
\end{equation}
i.e., when the acoustic wavelength $2\pi/k$, where $k = \sigma/c$, is smaller
than all length scales that characterize variations in $c$.  Under such
conditions the solution to (\ref{Helm}) can be written as a sum of terms,
each representing a locally plane wave,
\begin{equation}
\label{geom}
u({\bf r};\sigma) = \sum_j A_j({\bf r}) e^{i \sigma T_j({\bf r})}.
\end{equation}
Substitution of the geometric ansatz (\ref{geom}) into the Helmholtz equation
(\ref{Helm}) gives, after collecting terms in descending powers of $\sigma$,
the eikonal equation,
\begin{equation}
\label{eik}
(\nabla T)^2 = c^{-2},
\end{equation}
and the transport equation,
\begin{equation}
\label{trans}
\nabla(A^2 \nabla T) = 0.
\end{equation}
For notational simplicity we have dropped the subscript $j$ on $T$ and $A$ in
(\ref{eik}) and (\ref{trans}).
The solution to (\ref{eik}) can be reduced to the solution to the ray 
equations,
\begin{equation}
\label{ray2}
\frac{d{\bf r}}{d \tau} =  \frac{\partial {\cal H}}{\partial {\bf p}}, \;
\frac{d{\bf p}}{d \tau} = -\frac{\partial {\cal H}}{\partial {\bf r}},
\end{equation}
and
\begin{equation}
\label{dTdt}
\frac{dT}{d \tau} = {\cal L} = {\bf p} \cdot \frac{d{\bf r}}{d \tau} - {\cal H}
\end{equation}
where ${\bf p} = \nabla T$ is the ray slowness (also referred to as the
momentum) vector and
\begin{equation}
\label{H2}
{\cal H}({\bf p},{\bf r})
     = \frac{1}{2} \left( {\bf p}^2 - c^{-2}({\bf r}) \right) = 0.
\end{equation}
The independent (time-like) variable $\tau$ satisfies $d\tau/dl = c$ where
$dl = |d{\bf r}|$, and $d{\cal H}/d{\tau} = 0$.  The sum in (\ref{geom}) is
over all ray paths $z(r)$ that connect the source at $(z_0,0)$ and the receiver
at $(z,r)$.  The ray equations (\ref{ray2}-\ref{H2}) are seen to have
Hamiltonian form, which allows many well known results to be applied to the
acoustic problem in the short wave limit.  Equations (\ref{ray2}-\ref{H2})
describe the so-called optical-mechanical analogy of wave propagation in the
short wave limit.

For guided wave propagation in the direction of increasing $r$, the variable
$r$ can be used as the independent (time-like) variable and equations
(\ref{ray2} - \ref{H2}) may be rewritten
\begin{equation}
\label{ray1}
\frac{dz}{dr} =  \frac{\partial H}{\partial p}, \;
\frac{dp}{dr} = -\frac{\partial H}{\partial z},
\end{equation}
and
\begin{equation}
\label{dTdr}
\frac{dT}{dr} = L = p \frac{dz}{dr} - H
\end{equation}
where $p = \partial T/\partial z$ is the $z$-component of the slowness vector,
\begin{equation}
\label{H1}
H(p,z,r) = - \sqrt{c^{-2}(z,r) - p^2}
\end{equation}
is minus the $r$-component of the slowness vector, and
$dH/dr = \partial H/\partial r$.  These are the so-called one-way ray 
equations.
All subsequent analysis is based on these equations (or a parabolic
approximation to these equations, as described below), rather than the slightly
more general equations (\ref{ray2} - \ref{H2}).  Ray angles are defined by the
condition $dz/dr = \tan \varphi$ where $\varphi$ is measured relative to the
horizontal.  Using (\ref{ray1}) and (\ref{H1}) this reduces to
$cp = \sin \varphi$.  An immediate consequence of equations (\ref{ray1}),
independent of the form of $H(p,z,r)$, is
$\partial(dz/dr)/\partial z + \partial(dp/dr)/\partial p = 0$.  This is a
statement of Liouvilles theorem, expressing the incompressibility of flow
in phase space $(p, z)$.  The one-way ray equations (\ref{ray1}) have the same
form as the equations describing the motion of a relativistic particle with
mass $c^{-1}$ and Hamiltonian $H$.

The transport equation (\ref{trans}) can be reduced to a statement of
constancy of energy flux in ray tubes.  In a notation appropriate for use with
the one-way ray equations, the solution to the transport equation, assuming a
point source, for the $j$-th eigenray can be written
\begin{equation}
\label{gampl}
A_j(z,r) = A_{0j} \: |q_{21}|_j^{-1/2} \: e^{-i \mu_j \frac{\pi}{2}}.
\end{equation}
The matrix element $q_{21}$, defined below, describes the spreading of an
infinitesimal ray bundle.  At any fixed $r$, one has
\begin{equation}
\label{vareq}
\left (
\begin{array}{c}
\delta p \\
\delta z
\end{array}
\right )
= Q
\left (
\begin{array}{c}
\delta p_0 \\
\delta z_0
\end{array}
\right )
\; ,
\end{equation}
where the stability matrix
\begin{equation}
\label{Q}
Q =
\left (
\begin{array}{cc}
q_{11} & q_{12}\\ q_{21} & q_{22}
\end{array}
\right )
= \left (
\begin{array}{cc}
\left. \frac{\partial p}{\partial p_0} \right |_{z_0} &
\left. \frac{\partial p}{\partial z_0} \right |_{p_0}\\
\left. \frac{\partial z}{\partial p_0} \right |_{z_0} &
\left. \frac{\partial z}{\partial z_0} \right |_{p_0}
\end{array}
\right )
\; .
\end{equation}
Elements of this matrix evolve according to
\begin{equation}
\label{dQdr}
\frac{d}{dr}Q = KQ
\end{equation}
where $Q$ at $r = 0$ is the identity matrix, and
\begin{equation}
\label{K}
K =
\left (
\begin{array}{cc}
-{\partial ^2 H\over \partial z \partial p} &
-{\partial ^2 H\over \partial z^2} \\
{\partial ^2 H\over \partial p^2} &
{\partial ^2 H\over \partial z \partial p}
\end{array}
\right ).
\end{equation}
At caustics $q_{21}$ vanishes and the Maslov index $\mu$ advances by one
unit. (For waves propagating in three space dimensions, advances by two
units are possible.)  At these points diffractive corrections to
(\ref{gampl}) must be applied.  The normalization factor $A_{0j}$ is chosen
in such a way that close to the source (\ref{gampl}) matches the Green's
function for the corresponding wave equation, (\ref{Helm}) or a parabolic
approximation to the one-way form of this equation.  ($A_{0j}$ is different
for Helmholtz equation and parabolic equation rays; this small
difference is of no consequence in the discussion that follows.) 

\section*{\large B. The parabolic approximation}

The standard parabolic wave equation is
\begin{equation}
\label{PWE}
- \frac{i}{k_0} \frac{\partial \Psi}{\partial r} = \left(
      \frac{1}{2k_0^2} \frac{\partial^2}{\partial z^2} - c_0 U(z,r) \right) \Psi
\end{equation}
where $u(z,r) \approx \exp(ik_0r)\Psi(z,r)$, $k_0 = \sigma/c_0$,
\begin{equation}
\label{U}
U(z,r) = \frac{1}{2c_0} \left( 1 - \frac{c_0^2}{c^2(z,r)} \right),
\end{equation}
and $c_0$ is a reference sound speed.  The corresponding ray equations are
(\ref{ray1}) and (\ref{dTdr}) with $H(p,z,r)$ replaced by
\begin{equation}
\label{HPE}
H_{PE}(p,z,r) = \frac{c_0}{2}p^2 + U(z,r).
\end{equation}
Ray angles satisfy $c_0p = \tan \varphi$.  The parabolic approximation
is valid when ray angles are small and deviations of $c(z,r)$ from $c_0$
are small. The parabolic wave equation (\ref{PWE}) coincides with the
Schr\"{o}dinger equation with $r$ playing the role of time and $k_0^{-1}$
playing the role of Planck's constant $\hbar$.  It is useful to note this
analogy because many tools that have been developed to study quantum chaos,
discussed below, can be applied to the study of solutions of (\ref{PWE})
or (\ref{Helm}) under conditions in which the ray equations, (\ref{ray1})
or (\ref{ray2}), admit chaotic solutions.  In this regard it is noteworthy
that there is no quantum mechanical counterpart to transient sound fields
as these are characterized by the simultaneous presence of a continuum of
$k_0$ values.

\section*{\large C. Integrable ray systems}

It is well known that when the sound speed is a function of depth
only and the ocean boundaries are surfaces of constant $z$, the Helmholtz
equation (and the parabolic wave equation) admit separable solutions.  Under
the same conditions the ray equations also admit simple solutions that make
use of action-angle variables, and are said to be integrable.  The action-angle
formalism is important in the material in much of the remainder of the paper,
so the essential results are presented here.

When the sound speed is a function of depth only, ray trajectories are
periodic; the ray equations (\ref{ray1}) can be transformed, via a
canonical transformation, to a new set of ray equations in which the
Hamiltonian $H(I)$ is a function of the new momentum variable $I$ but is
independent of the new generalized coordinate $\theta$. In terms of the
action-angle variables $(I, \theta)$, ray trajectories are described by
the equations
\begin{equation}
\label{aa}
\frac{d\theta}{dr} = \frac{\partial H}{\partial I} \equiv \omega(I), \;
\frac{dI}{dr} = -\frac{\partial H}{\partial \theta} = 0.
\end{equation}
The solution to these equations is simply $I = \rm{constant}$,
$\theta(r) = \omega(I)r + \theta(0)$.  These equations (and their higher
dimensional counterparts) describe motion on a torus.  
The action variable can be written (see, e.g., Ref.
\cite{LLmech}) as a function of
$H$ (which is constant following each ray trajectory in a
range-independent environment),
\begin{equation}
\label{I}
I = \frac{1}{2\pi} \oint dz \, p(H,z)
   = \frac{1}{\pi} \int_{\check{z}(H)}^{\hat{z}(H)} dz \, p(H,z).
\end{equation}
The integration is over one cycle of the periodic ray trajectory, and at the
turning depths $c^{-1}(\check{z}(H)) = c^{-1}(\hat{z}(H)) = -H$.  The 
generating
function for the canonical transformation from $(p,z)$ to $(I,\theta)$ is
\begin{equation}
G(z,I) = \int^z dz^{\prime} \, p(H(I),z^{\prime})
\end{equation}
where the $H(I)$ can be obtained by inverting equation (\ref{I}), $p =
\partial G(z,I)/\partial z$, and
$\theta = \partial G(z,I)/\partial I$.

\section*{\large II. NONINTEGRABLE RAY SYSTEMS}

Realistic ocean structure is range-dependent, and the corresponding ray
systems are nonintegrable.  Thus, from a practical point of view, one is
forced to consider nonintegrable ray systems.  Also, from a more abstract
dynamical systems viewpoint, the property of integrability (separable
solutions) is known to be rather special.  It is
now customary to consider nonintegrable systems as being `generic'.  In
other words, if an `arbitrary' set of governing equations of motion were
considered, it would most likely be nonintegrable.  Only under rare
circumstances would a system be found to be integrable.  The other end of
the dynamical spectrum is the completely chaotic system which is discussed
in more detail ahead.  Another possibility includes near-integrable
dynamics, so-called because perturbation theory on the nearby integrable
system would be valid.  Finally, there is mixed dynamics in which both
regular and chaotic motion simultaneously exist in the system that are
accessed via different initial conditions.  Models of ray dynamics in
long-range ocean acoustic propagation tend to be in the mixed to fully
chaotic regimes.  

\section*{\large A. Nonlinear resonances and the KAM theorem}

Consider ray motion in an environment consisting of a range-independent sound
speed profile to which a small range-dependent perturbation is added.  Because
of the smallness of the sound speed perturbation, the perturbation to $H$ may
be assumed to be additive,
\begin{equation}
\label{H01}
H(p,z,r) = H_0(p,z) + \varepsilon H_1(p,z,r).
\end{equation}
For simplicity, first consider the case where $H_1$ is a periodic
function of $r$ with wavelength $\lambda = 2\pi/\Omega$.  It is well
known (see, e.g., Ref. \cite{AZ91}) that for this class of problems
canonical perturbation theory fails as nonlinear ray-medium resonances
are excited for those rays whose action values $I_0$ in the unperturbed
environment satisfy the condition
\begin{equation}
\label{res}
l \omega(I_0) = m \Omega
\end{equation}
for any pair of integers, $l$ and $m$.

A simple analysis (see, for example, Ref. \cite{AZ91}) shows that action
variables of rays captured into the resonance belong to the interval
\( I-\Delta I_{\max }<I<I+\Delta I_{\max } \)
with
\begin{equation}
\label{delI}
\Delta I_{\max }=2\sqrt{\varepsilon \bar{H}_1/|\omega ^{\prime }}|,
\end{equation}
where \( \omega^{\prime }=d\omega(I)/dI \) at \( I=I_0 \), and \(
\bar{H}_1 \) is a measure of the magnitude of $H_1$.  The quantity \(
\Delta I_{\max } \) represents the width of the resonance in terms of the
action variable.  The width of the resonance in terms of spatial
frequency can be approximately estimated as
\begin{equation}
\label{Om}
\Delta \omega = |\omega^{\prime }| \, \Delta I_{\max}/2
               = \sqrt{\varepsilon \bar{H}_1 \, |\omega^{\prime }|}.
\end{equation}
The phenomenon of nonlinear ray medium resonance plays an important role in
the emergence of ray chaos.  If there are at least two nonlinear resonances
centered at spatial frequencies \( \omega  \) and \( \omega +\delta \omega  \),
chaotic motion, according to Chirikov's criterion
(\cite{Chirikov71}-\cite{LL}), takes place when the condition
\begin{equation}
\label{Chirikov}
\frac{\Delta \omega }{\delta \omega }>1
\end{equation}
is satisfied, i.e., when the resonances overlap leading to the stochastic
instability of the system.

One might expect that, even for very small $\varepsilon$, all rays are
captured into a nearby resonance.  This turns out not to be the case.
According to the KAM theorem (see, e.g., Ref. \cite{Tabor}), for sufficiently
small $\varepsilon$ some of the tori of the unperturbed system are preserved
in the perturbed system, albeit in a slightly distorted form.  A condition
for the applicability of the KAM theorem is that the nondegeneracy condition
$\omega^{\prime} \neq 0$ must be satisfied.  This condition guarantees that
resonances are isolated provided $\varepsilon$ is sufficiently small.
Nondegeneracy violation will be discussed in more detail below. 

Because realistic sound speed structure in the ocean does not have
periodic range-dependence it is important to consider a larger class of
perturbation terms $H_1(p,z,r)$.  In Ref. \cite{B98} it is shown that the
KAM theorem applies to problems for which the perturbation term $H_1$
consists of a superposition of $N$ components, each of which is periodic
in range.  $N$ is assumed to be finite but is otherwise unrestricted.
The spatial periods need not be commensurable and there may be depth
structure associated with each periodic component.  Realistic ocean sound
speed structure -- internal-wave-induced perturbations \cite{CB}, for example --
can be described by a model of this type.  A consequence is that the mixture
of chaotic and regular trajectories (discussed below) that characterizes ray
motion in environments with periodic range-dependence applies to a much larger --
and more oceanographically realistic -- class of problems.

We note that there is a strong reason to study ray motion in periodic
environments even if they seem somewhat unrealistic.  In such environments
phase space structure can be observed by plotting pairs of points $(p,z)$
at integer multiples of the wavelength $\lambda$ of the perturbation.
Such a Poincar\'{e} map, which is found numerically, is a slice of the
ray motion in $(p,z,r \bmod \lambda)$.  In spite of the artificial nature
of the periodicity, the main features found are expected to be
qualitatively just like those found with aperiodic environments given
that the KAM theorem is equally valid whether the environment has
the property $c(z,r + \lambda) = c(z,r)$ for some $\lambda$ or not.

For some special problems ray dynamics and associated phase space
structures can be studied using an even simpler technique which
eliminates the need to numerically trace rays.  An example is described
in Ref. \cite{BTG}.  Here, ray motion in a bilinear model (constant sound
speed gradient above and below the sound channel axis) of the deep ocean
sound channel was considered.  It was shown that when the upper ocean
sound speed gradient oscillates periodically in range, successive
(separated by one ray cycle) iterates of axial ray angle and range satisfy
\begin{eqnarray}
\label{apm}
\phi_{n+1} & = & \phi_n + \varepsilon [\sin \rho_n
               + \sin (\rho_n + \phi_n + \varepsilon \sin \rho_n) ],
\nonumber \\
\rho_{n+1} & = & \rho_n + \phi_n + \varepsilon \sin \rho_n + \gamma
\phi_{n+1}.
\end{eqnarray}
Here $\varepsilon$ is the dimensionless perturbation strength, $\gamma$
is the ratio of the average upper ocean sound speed gradient $g$ to the
fixed lower ocean sound speed gradient, $\phi_n = (4\pi/g\lambda)
\varphi_n$ and $\rho_n= (2\pi/\lambda)r_n$.  These equations define an
area-preserving mapping, $\partial(\phi_{n+1},\rho_{n+1})/
\partial(\phi_n,\rho_n) = 1$; this condition is a discrete analog of
Liouville's theorem.  Because of their relative simplicity, area-preserving
mappings are widely used to study properties of nonintegrable Hamiltonian
systems.  Some of these properties will now be described. 

\section*{\large B. Ray chaos}

Figure 1(a) shows iterates of equation (\ref{apm}) for many sets of ray
initial conditions in an environment with moderately strong ($\varepsilon
= 0.15$) range-dependence.  Phase space is seen to consist of a mixture of
\begin{figure}[t]
\begin{center} 
\leavevmode 
\epsfxsize = 14.0cm 
\epsfbox{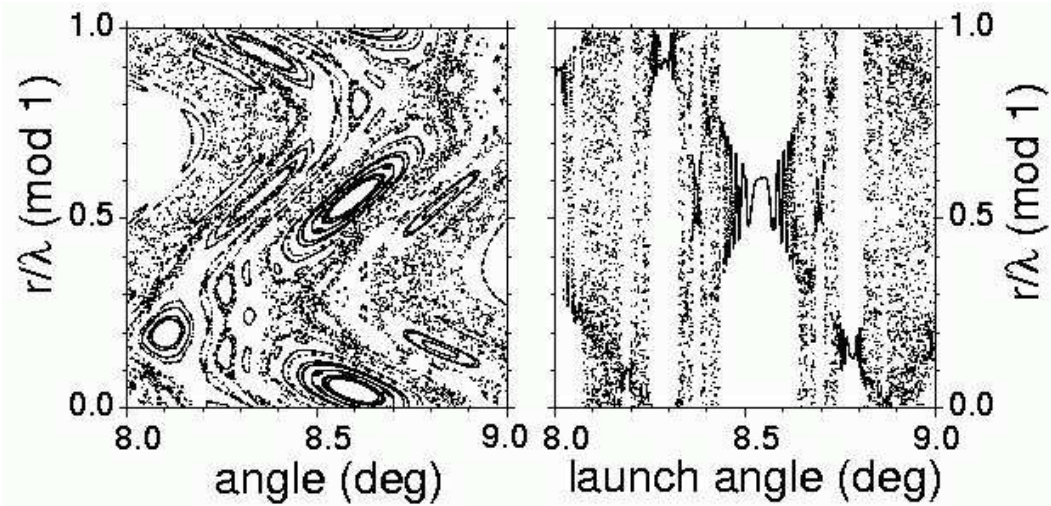} 
\end{center} 
\caption{Numerical simulations based on the area-preserving mapping (\ref{apm})
with $g = 1/(30 \; {\rm km})$, $\lambda = 10 \; {\rm km}$, $\gamma = 4$ and
$\varepsilon = 0.15$.  In both plots the ray initial conditions correspond to
those of an axial point source at $r = \lambda/2$.  Left panel: 500 iterates of
the mapping for 50 rays whose launch angles are uniformly distributed between
8 and 9 degrees.  Right panel: range after 250 ray cycles (each corresponding
to one iteration of the mapping) for 10,000 rays whose launch angles are
uniformly distributed between 8 and 9 degrees.  In most regions this sampling
interval is too large to resolve what should be a smooth function.} 
\label{figure1} 
\end{figure} 
regular and chaotic regions, often described as regular islands in a
chaotic sea.  For much smaller values of $\varepsilon$, chaotic regions
occupy only thin isolated bands of phase space.  Each such thin chaotic
band is associated with an isolated resonance.  As $\varepsilon$ is
increased, the widths of the resonances increase and nearby resonances
overlap leading to an intricate mixture of regular and chaotic regions,
as seen in Fig. 1(a).  Behavior of this type is typical of systems that
are constrained by KAM theory.  Figure 1(b) shows a plot of range vs.
launch angle after 250 ray cycles in the same environment, and using rays
emanating from the same fixed point that was used to produce Fig. 1(a).
The ray initial conditions used in both Figs. 1(a) and 1(b) fall on a
horizontal line (at $r = \lambda/2$) through the middle of Fig. 1(a).
Note that the islands that intersect this line in Fig. 1(a) are readily
identifiable in Fig. 1(b).  This observation is significant because plots
like Fig. 1(b) -- ray position vs. some continuous ray label -- can be
constructed in environments with nonperiodic range-dependence, providing
a simple means of identifying island-like structures.
At the boundaries of chaotic and regular domains -- in Fig. 1(a), for
example -- are usually invisible cantori.
These are Cantor-type invariant sets with fractal structure containing 
an infinity of holes.  Cantori act as partial barriers that inhibit the
diffusion of rays. Boundaries of chaotic regions contain small island
chains around which the density of points is very high.  A common
phenomenon is ``stickiness'' of island boundaries; after wandering in an
apparently random fashion in phase space, a chaotic trajectory may
approach a stable island, and stick to its border for some (possibly
long) time, during which it exhibits almost regular behavior
\cite{Edel}.  The presence of regular islands in phase space alters
the dispersion characteristics of the trajectories that lie in the
surrounding chaotic sea \cite{Edel}.  Details depend on the structure of
phase space but the phenomenon of anomalous diffusion (mean square
displacements of trajectories in phase space obeying scaling laws different
than that of a traditional random walk) seems to be generic.

An important property of chaotic trajectories is that they exhibit extreme
sensitivity, characterized by a positive Lyapunov exponent,
\begin{equation}
\label{Lyap}
\nu_L = \begin{array}{c} \lim \\ r\!\rightarrow\!\infty \end{array} \left(
  \frac{1}{r} \begin{array}{c} \lim \\ {\cal D}(0)\!\rightarrow\!0
\end{array}
       \ln \frac{{\cal D}(r)}{{\cal D}(0)} \right).
\end{equation}
Here ${\cal D}(r)$ is a measure of the separation between rays at range
$r$.   A closely related quantity is the Kolmogorov-Sinai entropy $h_{KS}$.
Loosely speaking, $h_{KS}$ is a measure of information increase following a
trajectory; a readable discussion of this topic can be found in
Ref.~\cite{Ott}.  For bounded dynamical systems $h_{KS} \sim \nu_L$ holds;
in open diffusive systems, their difference is proportional to a diffusion
constant~\cite{gaspard}.  A consequence of the extreme sensitivity of
chaotic rays is that the number of eigenrays connecting a fixed source
and receiver grows exponentially, like $\exp(h_{KS} r)$, on average in range
\cite{AZ85,SBT1,TT}. Also, the magnitudes of the variational quantities
$q_{ij}$ (see equation
\ref{Q}) can grow exponentially, on average, in range.  Reference
\cite{WTo} contains a detailed discussion of this topic.  A consequence
of this exponential growth is that the amplitudes of chaotic rays
(proportional to
$|\partial z(r)/\partial p(0)|^{-1/2}$ for (\ref{ray1}) or
$|\partial \rho_n/\partial \phi_0|^{-1/2}$ for (\ref{apm}), assuming a point
source) decay exponentially, on average, in range.  Note, however, that for
moderate to large $r$ or $n$ (measured in units of a typical value of
$\nu_L^{-1}$), plots of $\partial z(r)/\partial p(0)$ vs. $p(0)$ (see Fig. 5)
or $\partial \rho_n/\partial \phi_0$ vs. $\phi_0$ (see Fig. 1) have fractal-like
structure when both chaotic and regular trajectories are present.

Another consequence of extreme sensitivity of chaotic rays is that
deterministic predictions using finite precision numerical arithmetic is
limited to ranges less than some threshhold (which is proportional to
$N_b/\nu_L$ where $N_b$ is the number of bits used to specify the mantissa
of floating point numbers). This limitation on one's ability to make
deterministic predictions of chaotic ray trajectories at long range is tempered
by two related factors.  First, the shadowing lemma \cite{Ott} guarantees that,
for a large class of problems, numerically computed chaotic trajectories at
long range correspond to trajectories of rays whose initial conditions are
close to the specified values but are generally unknown.  Second, it is easy
to verify that statistical properties of discretely sampled distributions
of chaotic rays (occupying, for example, a small but finite area in phase space
at $r = 0$) evolve in a way that does not exhibit extreme sensitivity.

A somewhat stronger form of stability of distributions of chaotic rays is 
illustrated in Fig 2.  In phase space an aperture-limited compact source is
\begin{figure}[t] 
\begin{center} 
\epsfxsize = 13.0cm 
\epsfbox{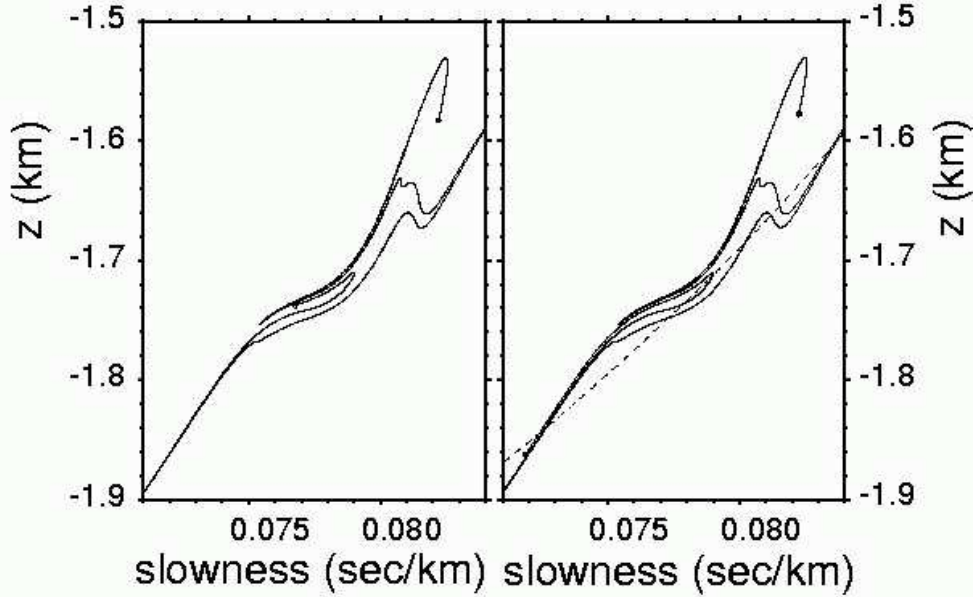} 
\end{center} 
\caption{Left panel: segment of a Lagrangian manifold for a fan of rays with
$z(0) = -1.100$ km, $8.575^o \leq \theta(0) \leq 8.625^o$ at $r = 500$ km
in the background environment shown in Fig. 3 with an internal-wave-induced
sound speed perturbation superimposed.  Right panel: segment of a
Lagrangian manifold for a fan of rays with $z(0) = -1.101$ km,
$8.575^o \leq \theta(0) \leq 8.625^o$ at $r = 500$ km in the same environment.
The endpoints of both Lagrangian manifolds are marked with small open circles.
In both panels, portions of the manifold segments, consisting of long thin
tendrils, extend beyond the plot boundaries.  The dashed curve in the right
panel is a portion of a surface $I$ = constant.} 
\label{figure2} 
\end{figure} 
represented as a line segment at constant depth $z = z_0$ bounded by
limiting values of $p_0$; this is an example of a Lagrangian manifold. 
Each point on such a manifold evolves in $r$ according to the ray
equations (\ref{ray1}). As a Lagrangian manifold evolves, it gets
stretched and folded, but does so without breaking or intersecting itself
(owing to phase space area preservation).  Fig. 2 illustrates one aspect
of a phenomenon that can be termed manifold stability.  In this figure
two Lagrangian manifolds with slightly different initial conditions are
plotted in phase space at a fixed range, $r = 500$ km.  Each manifold has
initial conditions $z = z_0$ (a constant), $8.575^o \leq \theta_0 \leq
8.625^o$.  For one of the manifolds
$z_0 = 1.100$ km; for the other $z_0 = 1.101$ km.  Both manifolds evolved in
the same environment, which is described below.  In this environment almost
all trajectories evolve chaotically, which leads to exponential growth in
range of the length of each manifold.  The combination of chaotic ray motion
and the small difference in initial manifold depth might lead one to expect
that the two manifolds should evolve very differently.  Figure 2 shows clearly,
however, that they have not.  This can be explained by noting that phase space
area is preserved.  Because of this constraint, the exponential stretching of
each manifold is balanced by an exponential contraction of phase space in the
transverse direction.  This contraction causes the two manifolds to get squeezed
closer to one another.  Note, however, that, in general, points on the two
manifolds with the same value of $\theta_0$ do not lie close to one another.
Loosely speaking, the two manifolds have slid relative to one another while
being stretched, folded and squeezed toward one another.  Trajectories with
nearby initial conditions will also be squeezed toward the same locus of
points.  Thus, although individual trajectories exhibit extreme sensitivity
under chaotic conditions, the associated exponential contraction of phase
space elements imparts a surprisingly strong form of stability on continuous
distributions of trajectories.  The impact of manifold stability on wavefield
evolution and stability is considered in Ref. \cite{CT}. 

The environment used to produce Fig. 2 is also used in subsequent numerical
work.  It consists of a range-independent background profile, shown in Fig. 3,
on which an internal-wave-induced sound speed perturbation field is
\begin{figure}[t] 
\begin{center} 
\leavevmode 
\epsfxsize = 6.0cm 
\epsfbox{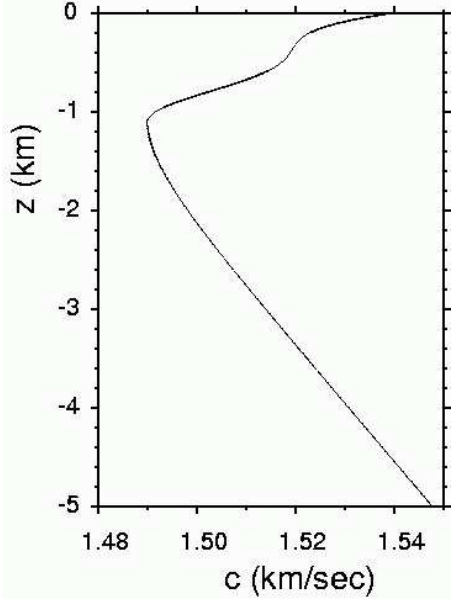} 
\end{center} 
\caption{Background sound speed profile used to produce Figs. 2, 4 and
5.} 
\label{figure3} 
\end{figure} 
superimposed.
The background profile is a Munk \cite{Mcan} profile modified in the upper
ocean, $c(z) = c_0(1 + \epsilon(\exp \eta - \eta - 1)) + c_u(z)$, with 
$c_0 = 1.49$ km/s, $\epsilon = 0.0057$, $\eta = (z - z_a)/B$, $B = 1$ km,
$z_a = -1.1$ km, and $c_u(z) = \delta \sin^2(\pi(z - z_a)/z_a)$ for $z > z_a$
with $\delta = 0.018$ km/s.  The internal-wave-induced sound speed perturbation
was computed using Eq. 19 of Ref. \cite{CB} with $y = t = 0$, i.e., a frozen
vertical slice of the internal wave field was assumed.  An exponential buoyancy
frequency $N(z) = N_0 \exp(z/B)$ (note that depths $z$ are negative with $z = 0$
at the sea surface) with $N_0 = 6$ cycles/hour was used.  The dimensionless
strength parameters $E$ and $\mu$ were taken to be $6.3 \times 10^{-5}$ and
$17.3$, respectively.  Numerically, a $2^{14}$ point FFT was used with
$\Delta k_x = 2\pi/1638.4$ km, $k_{x,max} = 2\pi/1$ km and $j_{max} = 30$.
It should be noted that this perturbation field is highly structured and fairly
realistically describes typical deep ocean environments.  Also, the assumed
background profile is similar to profiles found in much of the North Atlantic
Ocean.  

We have seen that the generic mixture of chaotic and nonchaotic ray
trajectories, the limited predictability of chaotic trajectories, and the
strong constraining influence of Liouville's theorem lead to very sublte
ray dynamics that blur the
distinction between traditional deterministic and stochastic viewpoints.
These and related ideas will be further explored in sections that follow.

\section*{\large C. Nondegeneracy violation}

Identifying conditions under which the KAM theorem does not apply is important
for two reasons.  First, inapplicability of KAM theory allows rays to behave
chaotically even for infinitesimally small perturbations to the background
sound speed structure.  Second, the character of this chaotic motion is
expected to differ from the chaotic motion of rays constrained by KAM theory.
This is because in systems constrained by KAM theory the presence of islands --
even small islands -- constrains the motion of nearby chaotic trajectories.
When KAM theory does not apply, the constraining influence of islands is lost.
In this subsection we briefly discuss a cause for KAM theory not to apply --
violation of the nondegeneracy condition.  We expect that
nondegeneracy violation is important in a wide variety of underwater acoustic
environments, but the topic has not yet been systematically explored. 

Ref. \cite{LL} makes a distinction between intrinsic and accidental
degeneracy.  In the former case $d\omega/dI$ vanishes for all trajectories.
Parabolic equation rays in the environment $(c_0/c(z))^2 = 1 - ((z-z_0)/h)^2$
and Helmholtz equation rays in the environment $c(z)/c_0 = \cosh((z-z_0)/h)$
are intrinsically degenerate.  These special problems are not realistic
oceanographically.  Also, it is worth noting that perturbations to intrinsically
degenerate systems often break up the degeneracy, leading to phase space
structure that is essentially the same as that seen in nondegenerate systems
\cite{LL}.  Accidental degeneracies, on the other hand, occur when
$d\omega/dI$ has isolated zeros -- when $d\omega/dI$ is plotted as a function
of launch angle for a point source, for example.  Accidental degeneracies
are structurally stable and are fairly common in realistic models of 
background ocean structure.  Thus we expect that accidental degeneracies
are important in ocean acoustics.  An example of an accidental degeneracy is
briefly discussed below.

The foregoing discussion can be quantified somewhat.  Recall that two
conditions for the validity of KAM theory are: 1) the dimensionless
perturbation strength in equation (\ref{H01}) $\varepsilon$ is small; and
2) the nondegeneracy condition,
\begin{equation}
\frac{d\omega}{dI} \neq 0
\label{eq4.1}
\end{equation}
is satisfied in the background environment, which is assumed to be
range-independent.  Zaslavsky \cite{Z98} has argued that these conditions may
be related; he argued that KAM theory is valid provided
\begin{equation}
\varepsilon < C |\alpha|^{\gamma}
\label{eq4.2}
\end{equation}
(for small $\varepsilon$ and $\alpha$) for some system-dependent $C$ and
$\gamma > 0$ where $\alpha$ is the dimensionless stability parameter
\begin{equation}
\alpha = \frac{I}{\omega} \frac{d\omega}{dI}.
\label{eq4.3}
\end{equation}
The term `weak chaos' has been used \cite{CSZ, TBG} to describe chaos that is
caused by an infinitesimal perturbation to an integrable system.  When $\alpha$
vanishes, equation (\ref{eq4.2}) will be violated for arbitrarily small nonzero
$\varepsilon$ and weak chaos is expected to be present.  In Ref. \cite{TBG} it
was demonstrated that rays in a small band surrounding the surface-grazing
ray exhibit weak chaos; for the surface grazing ray $d\omega/dI$ -- and
hence also $\alpha$ -- is undefined.

The stability parameter $\alpha$ can be computed from more familiar ray
quantities.  The angular frequency $\omega = 2\pi/R$ where $R$ is the ray cycle
distance.  For Helmholtz equation rays
\begin{equation}
R(p_r) = 2p_r \int_{\check{z}}^{\hat{z}} \frac{dz}{(c^{-2}(z) - p_r^2)^{1/2}}
\label{eq4.4}
\end{equation}
where $p_r$ is the $r$-component of the ray slowness vector, which is
constant following a ray trajectory in a range-independent environment.  The
upper turning depth of a ray $\hat{z}(p_r)$ satisfies
$c(\hat{z}(p_r)) = p_r^{-1}$, and similarly for the lower turning depth
$\check{z}(p_r)$.  Also, $cp_r = \cos \varphi$.  The action can be written
\begin{equation}
I(p_r) = \frac{1}{\pi}\int_{\check{z}}^{\hat{z}}dz \, (c^{-2}(z) - 
p_r^2)^{1/2}.
\label{eq4.5}
\end{equation}
Because $2 \pi dI/dp_r = -R(p_r)$, $\alpha = (2 \pi I/R^2)dR/dp_r$.  This
expression for $\alpha$ is convenient to evaluate numerically.

Figure 4 shows $\alpha$ as a function of axial ray angle in the environment
shown in Fig. 3.  It is seen that $\alpha$ has two zero crossings for rays
\begin{figure}[t] 
\begin{center} 
\leavevmode 
\epsfxsize = 12.0cm 
\epsfbox{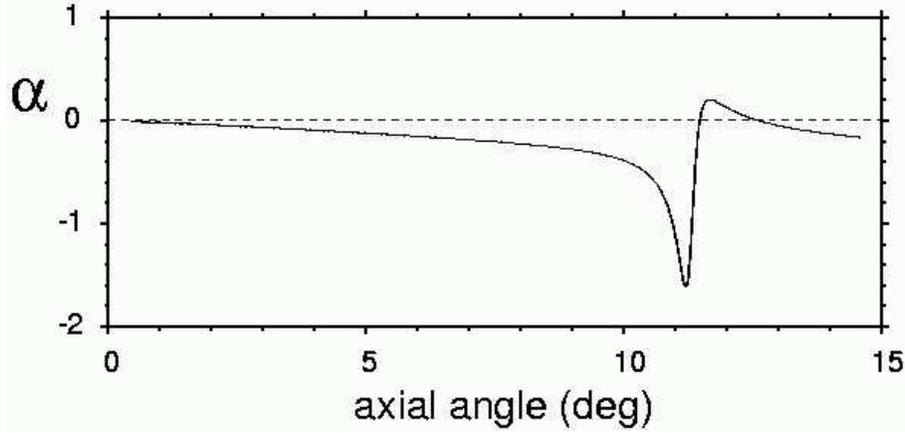} 
\end{center} 
\caption{Stability parameter $\alpha$ vs. axial ray angle in the environment
shown in Fig. 3. } 
\label{figure4} 
\end{figure} 
with axial angles near $12^o$.  Figure 5 shows a plot of ray depth vs. launch
angle for an axial source at a range of 1000 km in this environment with an
internal-wave-induced sound speed perturbation superimposed.  (The 
internal-wave-induced perturbation used to produce Fig. 5 is identical to that
used to produce Fig. 2 apart from a factor of two reduction in the strength
parameter $E$.)  In Fig. 5 it is seen
\begin{figure}[t] 
\begin{center} 
\leavevmode 
\epsfxsize = 10.0cm 
\epsfbox{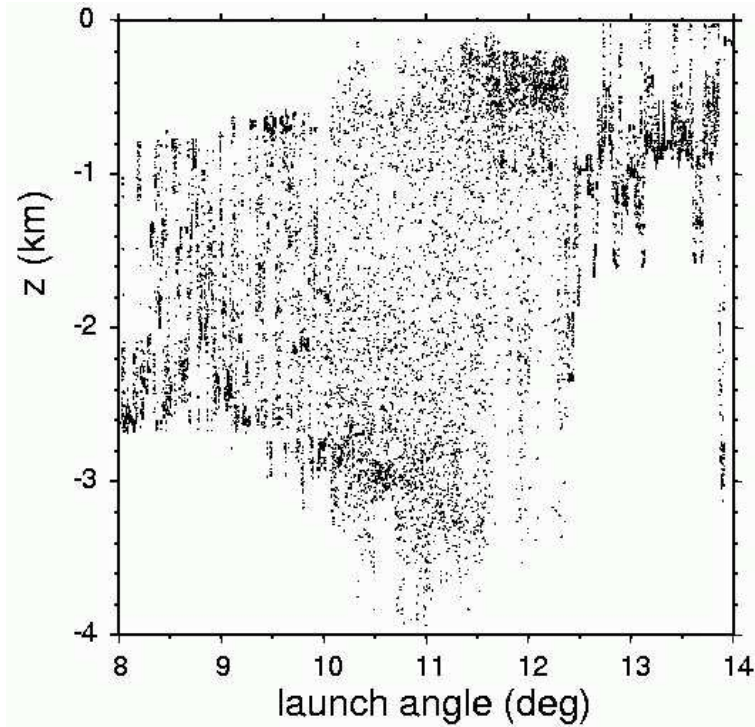} 
\end{center} 
\caption{Ray depth vs. launch angle for an axial source at a range of
1000 km in the background environment shown in Fig. 3 with an
internal-wave-induced sound speed perturbation superimposed.  In most
regions the sampling interval ($\Delta \theta_0 = 0.0005^o$) is too large
to resolve what should be a smooth function.} 
\label{figure5} 
\end{figure} 
that rays in the $10^o$ to $12^o$ band are evidently much less stable than
those outside this band.  Outside of this band there is evidence of apparently
nonchaotic island-like structures (compare to Fig. 1); inside this band there
is no indication -- at the resolution shown, at least -- of the presence of
such structures.  We attribute this difference to rays in the more stable
regions being constrained by KAM theory.  (The slight upward shift of the
unstable band of axial ray angles in Fig. 4 relative to Fig. 5 is not
unexpected.  In the presence of internal waves in the deep ocean, the upper
turning depths of rays whose axial angles are more than a few degrees tend to
move slightly toward the surface -- or equivalently, axial ray angles increase
slightly -- with increasing range.  This is because, following such a ray,
scattering is strongest near the ray's upper turning depth; any perturbation 
to the ray angle at the upper turning depth will make the ray steeper.) 

Because accidental degeneracies are not uncommon in realistic models of
(background) ocean structure, we expect the qualitative behavior exhibited in
Figs. 4 and 5 to be fairly common in ocean acoustics.  The occurrence of
accidental degeneracies  may be the cause of the finding of Simmen et al.
\cite{SFW} that, for identical sound speed perturbation fields, characteristics
of ray stability vs. launch angle curves depend strongly on the background sound
speed profile.  The connection between nondegeneracy violation and ray stability
needs to be further investigated.

\section*{\large III. RAY INTENSITY STATISTICS}

Ray intensity distributions are discussed in this section.  There are two
principal reasons for investigating this topic.  First, ray intensity
distributions are a useful diagnostic tool to study and quantify ray
dynamics, especially in environments with nonperiodic range-dependence where
Poincar\'{e} maps cannot be constructed.  Second, one expects -- on the
basis of the local plane wave expansion~(\ref{geom}) --
that a useful starting point for understanding wavefield intensity statistics
is understanding ray intensity statistics.  We emphasize that gaining an
understanding of ray intensity statistics is only the first step in this
process, as diffraction and interference effects must be accounted for in
making the transition to wavefield intensity statistics.  These complications
will be discussed in more detail below.

We shall confine our attention to a discussion of ray intensity statistics in a
simple idealized ocean model.  A more detailed account of the results presented
here can be found in~\cite{WTo}.  Some related results for the same idealized
problem are presented in~\cite{WTa}.  The model consists of an unbounded
homogeneous background on which an isotropic, single scale (described by a
Gaussian spectrum) random perturbation is superimposed.  It should not be
expected that all of the properties of ray intensity distributions in our
idealized problem carry over to long-range deep ocean propagation conditions,
which are characterized
by the presence of a background sound channel on which strongly inhomogeneous
and isotropic internal-wave-induced sound speed fluctuations with a power-law
spectrum are superimposed.  In spite of the idealized nature of the problem
considered here, there is ample justification for studying this problem.
First, this simple problem has surprisingly rich structure that must be
understood before more complicated problems can be successfully attacked.
Second, the results presented are expected to comprise a limiting case of those
that apply to more complicated problems.  And third, preliminary results
suggest that many of the results presented here apply generally to systems that
are far from integrable.  This topic will be discussed in more detail
elsewhere.

It is also worth mentioning that the results presented here are expected to
apply to other fields, as well.  Twinkling starlight is a familiar example,
but far more exotic systems with the same dynamical foundation exist.
Two such examples are the gravitational bending of light passing through galaxy
clusters~\cite{astro}, and the quantum mechanical waves associated with
low temperature, conduction electron transport through semiconductor
materials~\cite{lh94}.  Recent experiments of this latter system show the
electron transport breaking up into coherent channels that follow bundles
of classical rays that are only weakly unstable in spite of the
semiconductor acting as a random medium~\cite{topinka}.  This is a
manifestation of the remaining stable and nearly-stable rays mentioned in
Sect.~II in connection with the KAM theorem; such rays will show up again
later in this section.

Ray stability analysis relies on the stability matrix $Q$ defined in
Eq.~(\ref{vareq}) from which a great deal of information can be deduced.  It
describes the behavior of all the rays that remain within an infinitesimal
neighborhood of a reference ray over the course of its propagation.  As seen
in Eq.~(\ref{gampl}), geometric amplitudes for a point source are controlled
by the matrix element $q_{21}$; for a plane wave initial condition $q_{22}$ is
the relevant quantity.  For pedagogical purposes, however, it is simpler to
consider the trace of the matrix; for chaotic rays, its exponential rate of
increase and gross statistical properties are the same as those of any of the
$Q$ matrix elements.  Since $Q$ is diagonalizable by a linear, similarity
transformation
\begin{equation}
\label{diag}
\Lambda=L Q L^{-1}\Longrightarrow \left(\begin{array}{cc}
\lambda & 0 \\ 0 & \lambda^{-1}\end{array}\right) \; ,
\end{equation}
its trace is an invariant equal to the sum of the eigenvalues.  The last
form applies to systems with a single degree of freedom because the
determinant is unity.  Three distinct cases may arise.  The  first is
$|\mbox{Tr}(Q)|< 2$ which is linked to stable  motion, and it is
customary to denote $\lambda=\exp(i\theta r)$.  The second case is
$|\mbox{Tr}(Q)|=2$, and it is often called marginally stable because it
is the boundary case between stable and  unstable motion.  The third case
represents unstable motion, and is characterized by $|\mbox{Tr}(Q)|> 2$.
There it is customary to denote $\lambda=\pm \exp(\nu r)$  where $\nu$ is
positive and real.

In systems that are far from integrable, the vast majority of rays falls
into this last category.  Their largest Lyapunov exponent takes on an
alternative form to Eq.~(\ref{Lyap}), which can be expressed as
\begin{equation}
\label{lyapunov}
\nu_L \equiv \lim_{r\rightarrow\infty} {\ln\left(|\mbox{Tr}(Q)|
\right)\over r} \; .
\end{equation}
In the limit, $r\rightarrow\infty$, effectively all rays converge to a
unique value for $\nu_L$.  At finite ranges, each ray has different
properties, and it is quite helpful to introduce the concept of a finite
range Lyapunov exponent (sometimes known as stability exponent).  For unstable
orbits, $\mbox{Tr}(Q)=\lambda + 1/\lambda\approx \lambda$ with little
inaccuray giving
\begin{equation}
\label{nu}
\nu = {\ln|\mbox{Tr}(Q)|\over r} \; ;
\end{equation}
if necessary the more precise definition
$\nu= \mbox{cosh}^{-1}[\mbox{Tr}(Q)/2]/r$ could
be used, but it makes little difference for the discussion.  Thus, in the
limit, $\nu$ approaches $\nu_L$ for every ray.  The relationship between
these two exponents is rendered evident by introducing an ensemble of
potentials $U$ each of which generates a chaotic dynamical system of
fixed $\nu_L$.  The ensemble average of $\nu$ converges to $\nu_L$ without
the necessity of the $r\rightarrow \infty$ limit, except at very short
ranges.

A very important property of these relations is that even small or modest
fluctuations in $\nu$ produce immense fluctuations in $\mbox{Tr}(Q)$
which imply similar fluctuations in $q_{21}$.  Although, long propagation
ranges in chaotic systems imply vast numbers of eigenrays, the wave field
may be dominated by those far fewer terms that appear in the summation
with very small $q_{21}$.

An analytic expression for the root mean square exponential rate of
increase of $\mbox{Tr}(Q)$ has been derived using techniques relying on
Markovian assumptions~\cite{WTa}:
\begin{equation}
\label{trmeval}
\mbox{Tr}(Q)_{RMS}\sim\exp(\nu^\prime r)
\; .
\end{equation}
where
\begin{equation}
\label{nuprim}
\nu^\prime \approx \left({1 \over 2}\int_{0}^{\infty} d\xi
\left \langle \left . \frac{\partial^2 U(z;r-\xi)}{\partial z^2}
\right | _{z=z_0 \atop p=p_0} \left . \frac{\partial^2 U(z;r)}{\partial
z^2} \right | _{z=z_0 \atop p=p_0} \right \rangle \right)^{1/3}\; ,
\end{equation}
and the brackets, $<...>$, denote ensemble averaging over different
realizations of the potential $U$.  Numerical evaluation of
Eq.~(\ref{nuprim}) for model systems with some realism is usually
necessary, but analytic results are available for simplified models such
as Gaussian random single scale potentials.\\

Interestingly, $\nu^\prime$ is greater than the Lyapunov exponent,
independent of the range involved.  In fact, with increasing range it
rapidly approaches a constant.  The important distinction lies in whether
the ensemble averaging occurs before or after taking the natural
logarithm.  The fluctuations are strong enough that $\nu^\prime > \nu_L
\ (=\langle \nu \rangle)$ or alternatively
\begin{equation}
{\ln\langle|\mbox{Tr}(Q)|^2\rangle\over 2r} >
\left\langle{\ln|\mbox{Tr}(Q)|\over r} \right\rangle
\end{equation}
for all appreciable $r$.

It turns out that, for a homogeneous background and single scale Gaussian
random medium, to an excellent approximation except in the far tails,
the probability density of $\nu$ is a Gaussian of mean
$\nu_L$ and variance
\begin{equation}
\label{sigdep}
\sigma^2_\nu={\nu^\prime - \nu_L \over r} \; .
\end{equation}
This distribution has also been observed in numerical simulations based on
more realistic ocean models.  This topic will be discussed in more detail
elsewhere.  Numerical calculations also
suggest that the Gaussian statistics are obtained for each realization of
the random medium just by choosing different initial conditions; i.e. an
ensemble of media is not necessary.  A Gaussian density for $\nu$ implies
a lognormal density for $|\mbox{Tr}(Q)|$ whose parameters are fixed by the
Gaussian density's mean and variance.  A property of lognormal densities is
that any power of the variable also has a lognormal density.  Thus,
$|\mbox{Tr}(Q)|^\gamma$ has a lognormal density as well; $\gamma=-1/2$
relates to the semiclassical approximation.  However note that depending
on the value of $\gamma$, the lognormal density may fail as an approximation
less far out in the tails.

In the limit of long range, the fluctuations in $|\mbox{Tr}(Q)|$ grow
without bound despite $\nu$ approaching $\nu_L$ for every ray.  
Just as there are highly unstable rays, there are also rays
which are stable or nearly stable.  The lognormal density gives a
prediction for what approximate proportion is left for a given
propagation range.  Asymptotically, the proportion of nearly stable rays,
whose stability exponent is less than some small value $\nu_c$, decreases
exponentially with range as $(a_0/4 \pi r)^{1/2} \exp(-r/a_0)$ where
$a_0= 2(\nu^\prime -\nu_L)/(\nu_L - \nu_c)^2$.

The connection to the statistical behavior of wave field intensities arising
from the lognormal density of classical ray intensities is not yet understood.
There are a number of subtleties.  To begin with, for long ranges of
propagation, the most naive picture of semiclassical theory leads to the
expectation that the wavefield is made up of an extremely large number of
extremely small contributions.  On average, if diffusive growth in phase space
is neglected, the growth of the number of eigenrays and the decay of a typical
ray intensity (squared amplitude) should occur at the same exponential rate.
If this is true, energy conservation considerations dictate that at long range
the constituent ray arrivals have effectively random phases.  In other words,
a set of $N$ random, uncorrelated numbers of scale $N^{-1/2}$ maintains an
order unity summation as $N\rightarrow\infty$.  Chaotic dynamics is
deterministic, and at best, the semiclassical phases generated by the
dynamics are pseudo-random at long range.  For shorter ranges,
correlations amongst the magnitudes and phases could alter the
statistical predictions.  These correlations remain to be studied.
Furthermore, the possibility mentioned earlier that one or few very
strong terms at short range could dominate the sum increases the
difficulty in finding a unique approach to an asymptotic statistical
limit.

A second difficulty is that the lognormal distribution has long tails
indicative of its broad range of fluctuations.  It has the form
$y^{-a_0\ln y}$ which can be verified to approach zero for large or small
$y$ faster than any power of $y$, but does not decay exponentially.  A sum
of random numbers chosen from long tailed densities may not approach a
standard central theorem limit (the Lorentzian density remains Lorentzian
under repeated convolution), or may approach it very slowly.  For the ray
chaos problem, as more and more eigenrays exist with increasing range,
the breadth of the density is also increasing.  If the approach to a
central limit theorem is slow enough, then it might never be reached
under these circumstances; the limiting density still needs to be worked
out.

A third, extremely important difficulty is the appearance of caustics.
Their number proliferates at the same exponential rate for chaotic
systems as the number of eigenrays.  At the caustics, the semiclassical
expressions diverge and introduce infinities.  They correspond to the
extremely small values of instabilities in the tail of the lognormal
density where breakdown of the statistical laws are likely.  Thus, the
lognormal expression does not, for example, capture the physics of
singularity dominated fluctuations that are characterized by twinkling
exponents which depend on the types of caustics
present~\cite{twink,twink2,twink3}.  Some incorporation of deviations from
lognormality appears to be inescapable.  In addition, the presence of
exponentially proliferating numbers of caustics calls into question the
very relevance of semiclassical methods and their usefulness in predicting
the statistical properties of the wave field.  This consideration is
intimately linked with discussions of the validity of the semiclassical
approximation for chaotic systems of which we give an overview in the
next section.

A final noteworthy complication arises in the analysis of sound fields
produced by a broadband source.  Interference must still be accounted for,
but only among ray arrivals whose travel times fall within intervals whose
duration is the recriprocal bandwidth $(\Delta f)^{-1}$ of the source.  This
phenomenon is further complicated when pulse shapes and phase shifts at
caustics -- corresponding to Hilbert transforms in the time domain -- are
taken into account.  Surprisingly, in spite of the widespread use of broadband
sources in ocean acoustic experiments, we are unaware of any attempt to
systematically account for all of these complexities in a theory of broadband
intensity statistics.

\section*{\large IV. WAVE CHAOS}

`Wave chaos' is the study of wave systems that, in the ray limit, exhibit
unstable dynamics (i.e.~exponential divergence of neighboring rays);
therefore the ocean acoustics problem can be thought of as a wave chaos
problem.  A completely analogous definition of the phrase `quantum chaos'
is also widely used.  These two subjects are similar enough that much, if
not most, of the progress in either domain carries readily over into the
other; we therefore do not bother to distinguish them here.  It turns out
that there are significant conceptual difficulties with attempting to
associate wave chaos with an unbounded, exponential growth of wavefield
complexity in the hope that a straightforward generalization of the
underlying ray chaos manifests itself.  The shorthand explanation of this
difficulty is often crudely stated something like, ``the finite wavevector
(nonzero Planck's constant $\hbar$) smooths over the intricate, fine scale
details of the chaotic dynamics, not allowing them to be seen.''  The
correspondence principle for chaotic systems, in fact, is quite subtle,
and we avoid going into this fascinating subject except for the issue of
the breakdown of semiclassical theories which we summarize
next~\cite{ht}.  For further details though, we refer the reader to
Ref.~\cite{Bala} for a discussion of some aspects of wave chaos in
underwater acoustics, and to some recent literature relating to quantum
chaos~\cite{Chirikov,lh}.

The aspect of the breakdown of semiclassical theories which interests us
most for the purposes of this paper is whether or not it is possible to
construct, on solid theoretical ground, a ray-based theory valid beyond
where ray chaos has fully developed.  If not, ray-based predictions under 
chaotic conditions that match data in long-range propagation experiments would
have to be considered accidental, and not an explanation of the essential
physics of the problem.  It may turn out that the eventual answer to
this question is not unique, and depends upon which quantity one wishes to
explain.  For example, statistical predictions may be more robust than
detailed, point-wise, wavefield predictions.  On the other hand, if the
breakdown occurs on a scale much longer than that of the development of
ray chaos, there remains a great deal of theoretical work to be
pursued.  

There is some hope for optimism in this regard.  We begin by distinguishing
between two classes of dynamical systems.  The first class is that of simple,
chaotic systems.  Equations (\ref{apm}) define a system of this type.  The
equations of motion may be deceptively simple to write down, yet the solutions
highly complicated, and, for all intents and purposes, analytically intractable
(chaotic).  The second class has complicated equations in the sense that
the medium satisfies many of the criteria of randomness even though it is
taken here to be deterministic (we may not know which determinisitic
realization is given in a particular case).  One cannot take for
granted the equivalence of the properties of these two classes of
systems (chaotic versus random media), but we note that in certain limits
there exist a number of common results (such as exponential ray
instability with respect to initial conditions).  

In simple chaotic systems, about which more is known concerning
semiclassical breakdown, we have to be careful to distinguish three levels
of dynamics: classical, quantum, and semiclassical.  We emphasize that the
latter should be distinguished from the classical in that it takes
classical information as input, but it actually generates an approximate
construction of the quantum dynamics at the level of wave amplitudes
and phases.  The distinctions between time-evolving classical and quantum
expectation values of operators or classical and quantum probability
densities have been studied since the development of quantum mechanics.
Those quantities that correspond to each other initially are known to
propagate similarly before diverging over a finite time scale called the
Ehrenfest time.  For simplicity, we shall focus only on the fact that one
cannot delay the onset of interference phenomena in the quantum dynamics
beyond the Ehrenfest time, and interference is necessarily excluded from
the classical dynamics.  For chaotic systems, the Ehrenfest time depends
logarithmically on $\hbar$~\cite{GZ}, and beyond this time scale one
finds all manner of complications such as an exponential proliferation of
rays, increasing uncertainty whether the rays can even be calculated, and
proliferating caustics in semiclassical theories.   There is no debate
whether the classical and quantum dynamics diverge beyond the Ehrenfest
time (they do by definition); the relevant debate centers upon whether a
ray-based, semiclassical theory can surmount these difficulties.  

There are various `flavors' of semiclassical approximations possible.  For
example, the Wigner-Weyl calculus or other constructions of pseudo-phase
spaces~\cite{BJ} are ideally suited for exhibiting how the classical
limit emerges from quantum mechanics in a semiclassical limit, but they
are poorly adapted for describing interference phenomena.  Indeed,
mathematical proofs exist that such phase-space-based semiclassical
approximations cannot be extended beyond a logarithmic time scale;
see~\cite{BGP}.  In this scenario, it can be fruitful to examine smoothed
features of the wave field (as opposed to pointwise comparisons), such as can
be obtained through the ray-based construction of the coarse-grained Wigner
function~\cite{VZ2}.  However, the development of semiclassical approaches
that can be roughly described as time-dependent WKB theory handle
interference naturally though multiple stationary phase (saddle point)
contributions.  This approach has been applied to a number of paradigms
of chaos (the bakers map, the stadium, and the kicked rotor) with
excellent, numerical results extending well beyond the logarithmic time
scale~\cite{OTH,STH,TH}.  Arguments leading to the expectation that the
breakdown time depends algebraically on $\hbar$ have also been presented
~\cite{STH,TH,CL}.  

Semiclassical breakdown in an idealized, but highly structured, ocean model
consisting of a single realization of a ensemble with prescribed statistics
has been investigated by Brown and Wolfson.  They constructed the
full semiclassical approximation using the classical dynamics and a
Maslov-Chapman uniformization procedure.  In their comparisons, they
found the semiclassical approximation appeared to be working quite well
beyond the onset of ray chaos~\cite{BW}.  This work and those quoted for
simple chaotic systems are claiming that it is possible to extend a
semiclassical theory for both chaotic and random media problems beyond a
logarithmic time (range) scale, and thus, that ray-based semiclassical
theories are viable candidates purporting to explain the essential physics
of wave chaotic dynamical problems.  

\section*{\large V. CHAOS AND MODE COUPLING}

The connection between ray and modal expansions of acoustic wavefields in
range-independent environments is well understood (see, e.g., Refs.
\cite{Felsen81}-\cite{BrLys}) and some generalizations have been derived
\cite{Vir97}-\cite{Vir91} for range-dependent environments.  Thus,
it should come as no surprise that there is a quantifiable connection between
ray chaos and the modal description of the wavefield.  A detailed description
of this connection is described in Refs. \cite{VZ1} and \cite{V2000}.  In this
section, the connection between ray chaos and mode
coupling is briefly described.  To simplify the presentation somewhat, we make
use of a WKB analysis of the parabolic wave equation.  Consistent with our use
of the parabolic wave equation, the variables $p^{\prime} = c_0p$,
$H^{\prime} = c_0H_{PE}$, $U^{\prime} = c_0U$ and $I^{\prime} = c_0I$ are used
in this section after dropping the primes.  Also, $S = c_0T$ is used in place
of $T$. 

The principal result of this section is an approximate analytical expression
for mode amplitudes in term of ray-like quantities.
We discuss how typical features of ray chaos, such as exponential growth
(with range) of eigenrays contributing to the field point and coexistence of
chaotic and regular ray trajectories manifest themselves in the mode amplitude
range-dependence.  The phenomenon of nonlinear
ray-medium resonance that plays a crucial role in the emergence of ray chaos
is shown to have an analog for modes which we call the mode-medium resonance
\cite{VZ1}.  According to the heuristic criterion proposed by Chirikov
\cite{Chirikov71}-\cite{LL}, chaos is a result of an overlap of different
resonances. In terms of normal modes, chaos is shown to result from overlapping
mode-medium resonances leading to complicated and irregular range variations
of the modal structure. 

The normal mode representation of the solution to the parabolic wave equation
(\ref{PWE}) is obtained by expanding \( \Psi(z,r) \) in a sum of
eigenfunctions of the unperturbed Sturm-Liouville eigenvalue problem
\cite{LLquan,Porter},
\begin{equation}
\label{eig-eq}
-\frac{1}{2}\frac{d^{2}\psi_m}{dz^{2}}+k_0^2U_0(z)\psi_m
= k_0^2E_m\psi_m,
\end{equation}
where \( \psi_m \) and \( E_m \) are the eigenfunctions and
eigenvalues, respectively:
\begin{equation}
\label{modes}
\Psi(z,r)=\sum _mB_m(r)\, \psi_m(z).
\end{equation}
Here it is assumed that $U(z,r) = U_0(z) + \varepsilon V(z,r)$ and $\bar{V}$
is a measure of the magnitude of $V$.  Each term in the sum (\ref{modes})
represents a contribution from an individual normal mode.
In order to get simple semiclassical expressions for the amplitudes, \(B_m\),
we project the ray representation of $\Psi(z,r)$ of the form of Eq. 
(\ref{geom})
onto the normal modes.  Assuming the latter to be normalized in such a way that
\begin{equation}
\label{norm}
\int_{-\infty}^{\infty}dz\, \psi_m(z)\psi_n(z)=\delta_{mn},
\end{equation}
we need to evaluate the integrals
\begin{equation}
\label{Bm}
B_m(r) = \int_{-\infty}^{\infty}dz\, \Psi(z,r)\, \psi_m(z).
\end{equation}
Since we consider the geometric approximation to \( \Psi(z,r) \),
it is natural to use the same approximation for \( \psi_m(z) \). The
corresponding formulas are usually referred to as the WKB approximations to
the eigenfunctions \cite{BrLys,LLquan,Porter}. 

We shall assume that the potential \( U_0(z) \) is smooth, has only one
minimum and its walls tend to infinity as \( z\rightarrow \pm \infty  \).  When
this assumption is combined with use of the WKB approximation, the eigenvalues
of the action variable \( I_m \) are determined by the quantization rule
\begin{equation}
\label{quant}
k_0 I_m = m + \frac{1}{2}.
\end{equation}
Then the eigenvalues of the ``energy'' are given by the relation
\( E_m=E(I_m) \),
where the function \( E(I) \) is determined by Eq. (\ref{I}). In the same
approximation the \( m \)-th eigenfunction \( \psi_m(z) \) between its
turning points can be represented as \cite{BrLys,Porter}
\begin{equation}
\label{eigC}
\psi_m(z)=\psi_m^+(z)+\psi_m^-(z),
\end{equation}
with
\begin{equation}
\label{eigenC}
\psi_m^{\pm}(z)=Q_{m}(z)\, \exp \left[ \pm i\, \left( k_0 \int_{\check{z}}^z
  dz^{\prime}\,\sqrt{2[E_{m}-U_{0}(z^{\prime})]}-\frac{\pi}{4} \right) \right],
\end{equation}
\begin{equation}
Q_{m}(z)=\frac{1}{\sqrt{R_{m}}\left[ 2(E_{m}-U_{0}(z)\right] ^{1/4}},
\end{equation}
where \( R_{m} \) is the cycle length of the ray in the unperturbed waveguide
with \( E=E_{m} \). In this form the eigenfunction is represented as a sum
of two terms with rapidly varying phases and slowly varying amplitudes.
Substituting Eqs. (\ref{eigC}) and the ray representation of $\Psi(z,r)$ of
the form (\ref{geom}) into Eq. (\ref{Bm}) yields
a sum of integrals which can be approximately evaluated by applying the
stationary phase technique \cite{BrLys} . This has been done in Ref. 
\cite{VZ1}.
Earlier, a related result was obtained in Ref. \cite{Berman79}.  Here we
present only the final expressions for \( B_{m} \). 

It turns out that the amplitude of the \( m \)-th mode at the given range
\( r \) is formed by contributions from the rays with action variables equal
to that of the given mode, i.e., with
\begin{equation}
\label{staphm}
I=I_{m}.
\end{equation}
Equation (\ref{staphm}) singles out a set of rays which we shall call the
eigenrays for the \( m \)-th mode. The action variable on the left is 
considered
as a function of range, and initial values of the momentum \( p_{0} \) and the
coordinate \( z_{0} \), i.e.,
\begin{equation}
\label{I00}
I=I(p_0,z_0,r).
\end{equation}
This function is determined by the solutions to Eqs. (\ref{ray1}) and
(\ref{HPE}), and Eq. (\ref{I}).  For a point source the value of \( z_0 \) is
the same for all rays; substitution of Eq.(\ref{I00}) into Eq.(\ref{staphm})
then yields an equation for \( p_0 \), whose solutions define the initial
momenta of the eigenrays contributing to the \( m \)-th mode.

The process of identifying modal eigenrays is illustrated in the right
panel of Fig. 2.  Two curves are plotted: a surface of constant $I$ and a
segment of a Lagrangian manifold.  The Lagrangian manifold corresponds,
at $r = 0$, to a small angular aperture point source.  Each intersection
of the two curves corresponds to a modal eigenray.  Eleven such
intersections are shown.  Under chaotic conditions this number can grow
exponentially in range.  (In contrast, in all range-independent
environments there are two modal eigenrays for each mode, independent of
range, whose launch angles have opposite sign.) 

The mode amplitude is given by the sum
\begin{equation}
\label{ampo}
B_{m}=\sum _{n}\frac{1}
          {\sqrt{k_0\left| \partial I/\partial p_0\right| _{p_0=p_{0n}}}}\,
           e^{\, ik_0\Phi _{n}+i\beta_n},
\end{equation}
where each term represents a contribution from an eigenray with an initial
momentum \( p_{0n} \). The explicit expressions for the phase terms are (since
the subsequent formulas describe characteristics of a single eigenray, the
subscript \( n \) is omitted)
\begin{equation}
\label{phase1}
\Phi = S - S_{0}(z,I_{m}) \, \mbox{sgn}(p)
\end{equation}
and
\begin{equation}
\label{alpha1}
\beta = \left( \mbox {sgn}\left( \frac{\partial p/\partial p_0}
                                       {\partial z/\partial p_0}\right)
        - \mbox{sgn}(p) - 2\mu \right) \frac{\pi}{4}.
\end{equation}
Here \( z \) is the depth and \( p \) is the momentum of the eigenray
at a range \( r \), \( S \) and \( \mu  \) are its eikonal and Maslov index,
respectively. \\

Equations (\ref{I00})-(\ref{alpha1}) provide the analytical description of mode
amplitudes in a range-dependent environment through the parameters of ray
trajectories, i.e. through solutions to the Hamilton equations (\ref{ray1}).
These equations reduce the mode amplitude evaluation to a procedure quite
analogous to that generally used when evaluating the field amplitude at the
given point.  This involves solving the Hamilton (ray) equations, finding the
eigenrays, and calculating ray eikonals and some derivatives with respect to
initial values of ray parameters.
An important point should be stressed. Although we expand the wavefield over
eigenfunctions of the unperturbed waveguide, smallness of the perturbation has
not been assumed. Our small parameter is the acoustic wavelength, \( 
2\pi /k \), that should be substantially smaller than any physical scale
in the problem.

Having the comparatively simple expressions relating the mode amplitudes to
rays, we can now discuss how the complicated ray trajectory dynamics reveals
itself in the mode amplitude variations. For simplicity, we restrict our
attention to a waveguide with a weak (\( \varepsilon \) is now considered as a
small parameter) periodic range dependence with a spatial period
\( 2\pi /\Omega \), and we consider the case when only one mode is excited
at \( r=0 \), i.e.
\begin{equation}
u(0,z) = \psi_m(z).
\end{equation}
Analysis of the ray structure for this type of distributed source (see Refs.
\cite{VZ1,V2000}) shows that for \textit{}all rays initial values of the action
variable \( I \) are equal to \( I_m \) . A situation which we call
\textit{mode-medium resonance}
occurs when the value of \( I_m \) satisfies Eq. (\ref{res}).  Due to the
resonance, at ranges of order of \( 1/\Delta \omega  \) there will appear a
bundle of rays with the action variables \( I \) in the interval
\( |I-I_{0}| < \Delta I_{\max } \).  This means (see Eqs.(\ref{staphm})) that
starting with such ranges, the \( m \)-th mode is split into a group of
\( 2M \) modes with
\begin{equation}
\label{split}
M = \Delta I_{\max}/k_0
   = 2\sqrt{\varepsilon \bar{V}/\left| \omega^{\prime}\right| }/k_0.
\end{equation}\\
This expression is the modal analog of the standard expression (Eq. \ref{delI})
for the width of a ray resonance.  In the case of overlapping modal resonances
it is natural to expect a further broadening of a group of modes.  Moreover, as
we discussed earlier, the overlapping of resonances causes the emergence of ray
chaos with exponential proliferation of eigenrays.  Under chaotic conditions
the number of eigenrays contributing to a given mode also grows exponentially
with range, giving rise to a very complicated range dependence of mode
amplitudes.  Numerical simulations presented in Refs. \cite{VZ1,V2000} support
these statements. 

It might be assumed that that exponential proliferation of eigenrays
contributing to a given mode leads to statistical independence of 
mode amplitude
fluctuations under chaotic conditions.  We expect, however, that the problem of
describing mode amplitudes is considerably more rich and complicated.  First,
it should be recalled that generically the phase space of a chaotic
Hamiltonian system contains both chaotic regions and ``stable islands'' formed
by regular periodic trajectories.  Some such regular rays will be eigenrays for
some modes.  Their contributions to modes cannot be considered as stochastic.
Thus, we expect that under chaotic conditions there will be modes with
amplitudes composed of two constituents: a chaotic one and a regular one.
Numerical results illustrating this statement have been presented in Ref.
\cite{V2000}.  Another important phenomenon typical of chaotic dynamics, which
may affect modal structure variations is stickiness, i.e., the presence of
segments of a chaotic trajectory which exhibit almost regular behavior.  The
interval over which apparently regular behavior is observed can be long.  In
principle, one can presume that stickiness may cause some long-lasting
correlations of mode amplitudes \cite{Bala}.

Our ray-based description of normal mode amplitudes has restrictions that are
very much like those of standard ray theory. In particular, at some points
the contribution from an eigenray to a given mode can be infinite.  This occurs
when the derivative in the denominator in Eq. (\ref{ampo}) vanishes.  (The same
comment applies to Eq. (\ref{Gm}), below.)  Such divergences represent analogs
of standard ray theory caustics.  Under conditions of ray chaos the number of
such caustics grows exponentially with range, spoiling applicability of the
ray-based description already at short distances.  This issue is discussed in
Ref. \cite{V2000}.  On the other hand, in Ref. \cite{V2000} (see also Ref.
\cite{VZ2}) it was demonstrated numerically that the approach considered in
this section can properly predict squared mode amplitudes smoothed over the
mode number at ranges of order of at least ten inverse Lyapunov exponents. This
result is rather encouraging.  It gives us hope that energy redistribution
between modes can be comparatively easily analyzed using simple ray 
calculations at ranges of the order of a few thousand km.

So far, we have considered only a cw field. For a signal with a finite
bandwidth, the mode sum must include an integration over frequency $\sigma$.
The pulse signal at the point \( (z,r) \) can be represented as
\begin{equation}
\label{ut}
u(z,r,t)=\sum _m u_m(z,r,t),
\end{equation}
where
\begin{equation}
\label{ut1}
u_m(z,r,t) = \int d\sigma \, s(\sigma)\, B_{m}(r,\sigma )\,
  \psi_m(z,\sigma)\, e^{i\sigma \left( \frac{r}{c_0}-t\right) }
\end{equation}
with \( s(\sigma) \) being the spectrum of the initially radiated pulse. In
Eq. (\ref{ut1}) we indicate explicitly the dependencies of \( B_m \) and
\( \psi_m \) on \( \sigma  \) which have been omitted until now.  Each term,
\( u_m(z,r,t) \), in the sum (\ref{ut}) can be interpreted as a pulse carried
by an individual mode and we shall call it the ``mode'' pulse.  The mode
pulse, in turn, can be regarded as a superposition of elementary pulses
representing contributions from different terms in the sum (\ref{ampo}):
\begin{equation}
\label{ut2}
u_m(z,r,t)=\sum _{n}\int d\sigma \, s(\sigma)\, G_{m}(z,r,\sigma)\,
                    e^{i\sigma \left( \frac{\Phi_n + r}{c_0}-t\right) },
\end{equation}
where
\begin{equation}
\label{Gm}
G_m(z,r,\sigma)=\frac{e^{i\beta_n}}
                {\sqrt{k_0r\left| \partial I/\partial p_0\right| _{p_0=p_{0n}}}}
                \frac{2\cos \left( k_0 \int_{\check{z}}^{z}dz^{\prime}\,
                \sqrt{2[E_m-U_0(z^{\prime})]}-\pi /4\right) }
                {\sqrt{R_{m}}\left[ 2(E_{m}-U_{0}(z)\right] ^{1/4}}.
\end{equation}
The above expressions depend on mode parameters with the subscript \( m \)
and eigenrays parameters with the subscript \( n \). Note that both types of
parameters depend on frequency. Although the trajectories of eigenrays
contributing to the given mode obey frequency-independent Hamilton equations
(\ref{ray1}), their starting momenta determined by Eq. (\ref{staphm}) depend on
the eigenvalue \( I_{m} \).  But the latter, according to Eq. (\ref{quant}),
does depend on frequency.  Under chaotic conditions the number of terms in
Eq. (\ref{ut2}) can be huge leading to a very complicated shape of the mode
pulse. 

The expectation that mode pulses are very complicated under chaotic conditions
is consistent with the numerical results in Ref. \cite{CF}, where broadband
parabolic-equation simulations of sound transmission through a deep water
acoustic waveguide with inhomogeneities induced by random internal waves are
described.  It was shown that, due to mode coupling, mode pulses were several
times longer than was the case in the background range-independent environment,
and acquired irregular shapes.  In contrast, from the ray perspective, the same
weak inhomogeneities caused steep eigenrays to split into clusters of eigenrays
(micromultipaths) whose travel time spreads were small and whose centroid had
a travel time that was close to that of the eigenray in the background
environment.  (Recall also section II of the present paper.)  The authors of
Ref. \cite{CF} concluded that ``while the high modes may be strongly 
affected by
internal waves they are coherent enough that when they are synthesized together
localized wave front results.''  A qualitative explanation of this phenomenon
has been offered in Ref. \cite{V99F}.  In that paper it was shown that mode
pulses may be considerably distorted due to mode coupling already at ranges so
short that chaotic ray dynamics has not yet had a chance to reveal itself and
every mode is formed by contributions from only two eigenrays.  It was also
demonstrated how distorted pulses carried by individual modes can combine to
produce much less distorted ray-like pulses at the receiver.

Although the mode coupling relations presented above are not easy
to test experimentally, we emphasize that these results are important because
of the insight they provide into the underlying propagation physics.  In this
regard, it should be noted that the mode coupling relations presented above
directly address the connection between ray chaos and finite frequency
propagation effects, i.e., the wave chaos problem.

\section*{\large VI. DISCUSSION}

In this paper we have reviewed results relating to ray dynamics in ocean
acoustics.  All of these results are intimately linked to the Hamiltonian
structure of the ray equations.  Most previous studies have emphasized the
applicability of KAM theory to oceans with periodic range-dependence and the
extreme sensitivity of individual chaotic ray trajectories.  To complement
these ideas, considerable attention has been focussed on oceans with
nonperiodic range-dependence, and we have discussed some forms of stability
of distributions of chaotic rays.  Also, we have discussed subjects such as
nondegeneracy violation, ray intensity statistics and the connection between
ray chaos and mode coupling that either haven't, or have only recently, been
explored in an underwater acoustic context.

Although our understanding of ray dynamics is currently incomplete, it should
be clear that the most pressing problem in this context is to better
understand the connection between the ray dynamics and the corresponding
finite frequency wavefields.  For example, even the seemingly simple task
of translating ray intensity statistics to wavefield intensity statistics is
complicated by the necessity of making an additional assumption about relative
phases and correcting for diffractive effects. 

It is our hope that the theoretical results presented in this paper provide a
foundation for the analysis of measurements of sound fields at long range in
the deep ocean.  An analysis of this type will be presented in a forthcoming
paper.  Our long-term goal of developing tools that can be used to assist in
the analysis of measurements accounts, in large part, for the considerable
attention that we have devoted to ocean structures with nonperiodic
range-dependence.

The ideas and results that we have discussed differ in some important respects
from more commonly applied ideas and results associated with the study of
wave propagation in random media (WPRM).  For example, most long-range
underwater acoustic propagation WPRM theories (see, e.g., \cite{FDMWZ}) assume
that stochasticity is caused exclusively by internal waves.  From the
deterministic chaos point of view, this assumption is difficult to justify
inasmuch as non-internal-wave (e.g., mesoscale) structure may excite
ray-medium resonances and chaos.  Also, in the deterministic chaos point of
view, the excitation of ray-medium resonances generally leads to a mixed
phase space.  Most approaches to WPRM, on the other hand, invoke the assumption
that the perturbation to the background sound speed structure is
delta-correlated; this leads to stochasticity, but not to a mixed phase space.
These topics will be discussed in more detail in future work.

Finally, we wish to remark that the importance of ray methods is not diminished
by recent advances, both theoretical and computational, in the development of
full wave models, such as those based on parabolic equations.  The latter are
indispensible computational tools in many applications.  In contrast, the
principal virtue of ray methods is that they provide insight into the
underlying wave physics that is difficult, if not impossible, to obtain by
any other means.  For this reason ray methods remain important.

\section*{\large ACKNOWLEDGMENTS}

We thank F. Tappert for the benefit of discussions on many of the topics
included in this paper.  This work was supported by Code 321 OA of the U.S.
Office of Naval Research.

\end{document}